\documentclass{emulateapj}

\shorttitle{Disk Assembly: Bulgeless Galaxies and M${}_{\ast}$-TF}
\shortauthors{Miller et al.}

\begin{document}

\title{Uncovering Drivers of Disk Assembly: Bulgeless Galaxies and \\ the Stellar Mass Tully-Fisher Relation}

\author{
Sarah H. Miller\altaffilmark{1,2,3},
Mark Sullivan\altaffilmark{4,1}, \&
Richard S. Ellis\altaffilmark{2}
}

\email{sarah@sarahholmesmiller.com} 

\altaffiltext{1}{Department of Physics (Astrophysics), University of Oxford, Keble Road, Oxford, OX1 3RH, UK}
\altaffiltext{2}{California Institute of Technology, Pasadena, CA 91125}
\altaffiltext{3}{University of California, Riverside, CA 92521}
\altaffiltext{4}{School of Physics and Astronomy, University of Southampton, Southampton, SO17 1BJ, UK}

\begin{abstract}
  In order to determine what processes govern the assembly history of
  galaxies with rotating disks, we examine the stellar mass
  Tully-Fisher relation over a wide range in redshift partitioned
  according to whether or not galaxies contain a prominent bulge.
  Using our earlier Keck spectroscopic sample, for which bulge/total
  parameters are available from analyses of HST images, we find that
  bulgeless disk galaxies with $z>0.8$ present a significant offset
  from the local Tully-Fisher relation whereas, at all redshifts
  probed, those with significant bulges fall along the local relation.
  Our results support the suggestion that bulge growth may somehow expedite the
  maturing of disk galaxies onto the Tully-Fisher relation. We discuss
  a variety of physical hypotheses that may explain this result in the
  context of kinematic observations of star-forming galaxies at
  redshifts $z=0$ and $z>2$.
\end{abstract}

\section{Introduction}

A major goal in galaxy evolution studies is to fundamentally
understand the evolving dynamical and morphological forms of galaxies
\citep{robert1969}. The favored method of tracking the assembly of
stellar mass as a fraction of the total mass in rotationally-supported
galaxies is the redshift-dependent Tully-Fisher (TF) relation, which
was first explored at $z\sim1$ by \citet{vogt1996,vogt1997}.
Subsequent studies of the stellar mass ($M_{\ast}$)-TF relation at
intermediate-to-high redshifts revealed scatters $\sim3\times$ larger
than that of the local relation
\citep[e.g.,][]{consel2005,kassin2007,vergan2012}. This increased
scatter was initially thought to represent a weaker coupling between
stellar and dynamical mass, precluding detailed studies of the
evolution in either slope or normalization. However, we showed in
\citet{miller2011} and \citet{miller2012a} (hereafter, M11 and M12,
respectively) with data of improved signal/noise and refined modeling techniques, that
the $M_{\ast}$-TF relation is actually well-established at $z\simeq1$
with a scatter comparable to that seen in the local relation.  Moreover,
in M12, we demonstrated that the relation holds for most disk galaxies
since $z\simeq1.7$, thereby posing a challenge of how to explain the
rapid evolution in kinematic behavior since $z\sim2$ where
star-forming galaxies are morphologically-irregular and dispersion
dominated~\citep{forste2006,law2007b,forste2009}. To the extent that a
TF relation can be examined at
$z\simeq2$~\citep{cresci2009,gneruc2011}, a normalization increase of
0.4 dex is seen over $\simeq$1 Gyr to $z\simeq$1.5, in contrast to
only 0.02 $\pm$0.02 dex over the subsequent 9 Gyr.

Since in M12 the TF scatter is observed to decline by 60\% from $z\simeq$1.7 to
$z\simeq$1, in this present paper we seek to examine whether this
might arise from physical properties governing the evolution onto the TF
relation. We target our attention on the morphological appearance of
each galaxy, specifically the bulge-to-total
ratio. Bulgeless disks representing at least $15\%$ of local galaxy populations \citep{kormen2010} provide an interesting challenge for hierarchical $\Lambda$CDM galaxy formation  \citep[which leads to inevitable bulge-growth without substantial feedback:][]{robert2006,govern2010}. We test whether the high redshift $M_{\ast}$-TF relation can be better understood when tracking the mature, bulge-dominated population of galaxies separately from the evolving population of bulgeless systems experiencing a more secular formation process.

Throughout the paper we adopt a \citet{chabri2003} initial mass
function and a $\Omega_{\Lambda}$ = 0.7, $\Omega_{m}$ = 0.3, $H_0$ =
70 km s$^{-1}$ Mpc$^{-1}$ cosmology. All magnitudes refer to those in
the AB system (Oke 1974).

\section{Dynamical Data and Stellar Masses} \label{sec:data}

The key measurements required to follow the evolving $M_{\ast}$-TF relations are disk kinematics as parameterized through rotation curve model fits, and stellar mass estimates 
derived from multi-band photometric data. Our earlier papers (M11, M12) describe the relevant data and their reduction in considerable detail so we provide only a brief summary here.

Our spectroscopic sample was selected from Hubble Space Telescope (HST) Advanced Camera for Surveys (ACS) imaging data in various survey fields complete to an apparent magnitude of $i$=22.5 and is morphology inclusive, containing irregular and merging systems as well as regular spirals with and without bulges. Keck spectroscopy was undertaken for 236 galaxies with 0.2 $<z<$ 1.3 at a median spectral resolution of 30 km s$^{-1}$
using the DEep Imaging Multi-Object Spectrograph\citep[DEIMOS][]{faber2003} and, subsequently, 70~~$1.0\lesssim z<1.7$ galaxies were targeted at a median resolution of 57~km~s${}^{-1}$ with the 
Low Resolution Imaging Spectrograph \citep[LRIS][]{oke1995} equipped with a red-sensitive CCD.  An unique aspect of both spectroscopic campaigns was the use of long exposure times (4-8 hrs) essential 
for tracking the rotation curves to the flattening radius (see M1 for details). Rotation curves were derived using various emission lines (H$\alpha$, [O\textsc{ii}], and [O\textsc{iii}] depending on
the galaxy redshift. As discussed in M11, we account for position-dependent dispersion and emission brightness profile, convolved with the seeing,  and adopt an arctangent function. We use inclination-corrected fiducial velocity measurements at 3.2 times the disk scale radius. The final sample for consideration here comprises 171 galaxies for which rotation curves could be determined (this is all galaxies except for spectrally compact or passive galaxies: see M11 \& M12 for details).

Stellar mass estimates are determined using a combination of ground-based $K$-band infrared imaging, multi-band optical photometry, and spectroscopic redshift information using the 
spectral energy distribution (SED) fitting technique first utilized by \citet{brinch2000}. Measured magnitudes in multiple bands were applied using a Bayesian code based on the precepts 
discussed in \citet{kauffm2003}, and later \citet{bundy2005}. Using probability distribution functions that incorporate uncertainties in the photometry, the stellar mass uncertainty is better 
than 0.2 dex for 83\% of our sample. 

\section{Morphological Data}

Our primary goal is to investigate the possible role bulge formation may play in the apparent rapid evolution of the $M_{\ast}$-TF relation from $z\simeq$2 to z$\simeq$1.
We facilitate this investigation with \textsc{galfit} \citep{peng2010}. As we required disk scale lengths for earlier applications,  the bulge-to-disk decomposition procedure described
is similar to that in M11, M12, and \citet{miller2012b} and so only briefly discuss the procedure here.

We run \textsc{galfit} on each galaxy 1000 times, varying the initial
parameters in gaussian distributions based on their SExtractor
\citep{ber96} values. For each object we attempt to fit a
deVaucouleurs bulge profile and an exponential disk component, where
the fit parameters are the center position, total magnitude
$m_{\mathrm{tot}}$, effective radius $R_e$ (scale radius, $r_s$, for
an exponential disk), S\'{e}rsic index $n$ (fixed to $n=4$ for
deVaucouleurs and $n=1$ for disk), axis ratio $q$, and position angle
$\phi$.  Where physical bulge solutions are not found, we re-fit the
galaxy with an index-varying single S\'{e}rsic component (indices
typically lie between $1<n<4$). Such cases generally represent disk
galaxies which are bulgeless and/or irregular. Disk sizes,
inclinations and position angles were taken from best-fit disk
components if more than one component was fit. Final parameter
uncertainties from the Monte Carlo distributions are better than 5\%
on average, and we add these uncertainties in quadrature to the
photometric errors from \textsc{galfit}. The scale radii, position
angles and inclinations are typically measured better than 10\%.
Uncertainties are propagated through to TF parameters, resulting in
larger errors for those galaxies which are difficult to constrain.

In the DEIMOS sample, $\sim$40\% were adequately fit using a 2-component decomposition, and $\sim$60\% benefitted from a single $n$-varying S\'{e}rsic profile fit. In the LRIS sample, the relevant percentages were $\sim$63\% and $\sim$37\%, respectively. This serves as a good indication of the morphological distribution of our sample;
less than half are well-formed spirals with a clear bulge (\S \ref{sec:data}). Where HST data is available in multiple bands we compared \textsc{galfit} runs between bands to test for differences 
in the scale radius determination as a function of redshift. The scale radii are consistent among the bands indicating no significant redshift-dependent bias (less than 5\% in the DEIMOS sample and $<$10\% for that of LRIS). In order to maximize signal/noise we use the \textsc{galfit} results from the reddest 
available filter (F814W or F850LP). 

A crucial issue affecting classification at high redshift is the `morphological k-correction' -- the change in apparent morphology with increasing redshift following the drift
blue ward in rest-frame wavelength. This is potentially troublesome for $z>1$ where the the F814W and F850LP images sample the younger star-forming regions
rather than the older, redder populations that dominate the stellar mass at lower redshift.  As such, there is a danger of underestimating the bulge contribution. 

The HST near infrared Wide Field Camera 3 (WFC3/IR) provides a F160W
filter, which at $1<z<2$ provides rest-frame optical light and is
therefore ideal for the bulge-to-disk decompositions we seek.  While
deep WFC3/IR F160W imaging from the CANDELS survey \citep{koekem2011} is
available for one-fifth of our sample, the majority of the combined
LRIS and DEIMOS samples are unfortunately in GOODS North (the WFC3/IR
coverage of which will not complete for at least another year). However,
for the purposes of this paper we seek only to demonstrate that use of
the ACS data to classify the sample broadly into bulge-less and bulge-dominated
subsets does not induce significant biases. As we discuss below, we will 
split our overall sample according to a dividing bulge-to-total ratio ($B/T$) = 0.1. 
With this division, we find, for the data with present WFC3/IR coverage, that 
morphological classifications into these two categories are consistently
made between the WFC3/IR and ACS data for 85\% of the total sample. 

\section{Results}

\begin{figure*}
   \centering
      \centering
   \includegraphics[width=4.5in]{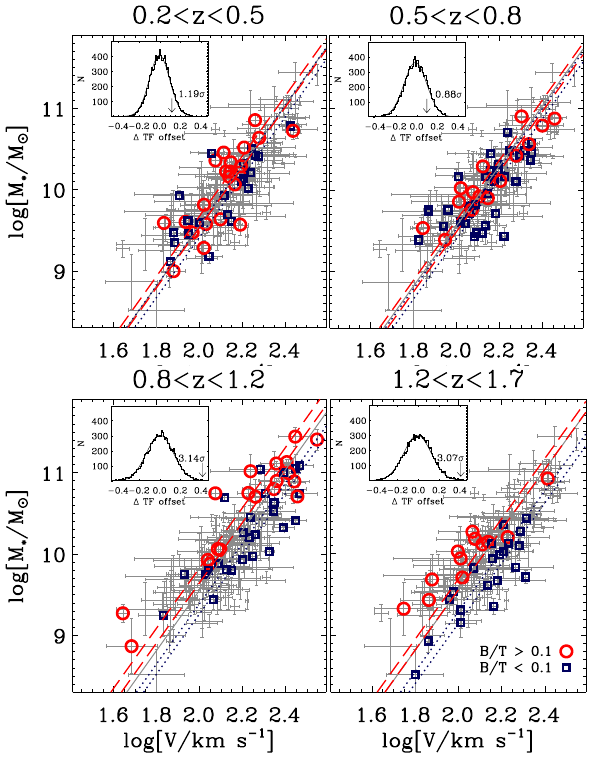}
      \caption{The $M_{\ast}$-TF relations in four redshift bins. Galaxies with bulges ($B/T$ $>$ 0.1) are marked with red circles, and bulgeless/irregular galaxies ($B/T$ $<$ 0.1) are marked with navy boxes. The relation of bulge galaxies (red dashed lines, 1-$\sigma$) lie along the local relation in each $z$ bin, whereas the bulgleless relation (navy dotted lines, 1-$\sigma$) are offset from the local relation in the two highest $z$-bins. To aid the eye, we plot the total sample with light grey error bars in each panel and include a solid grey line to mark the local relation.  To understand the significance of the offsets, a Monte-Carlo `bootstrap' analysis is conducted for each $z$-bin, plotted in the inset panels along with the location and significance of the true offset.}
   \label{fig:bulgeless_tf}
\end{figure*}

\begin{figure}
   \centering
   \includegraphics[width=3in]{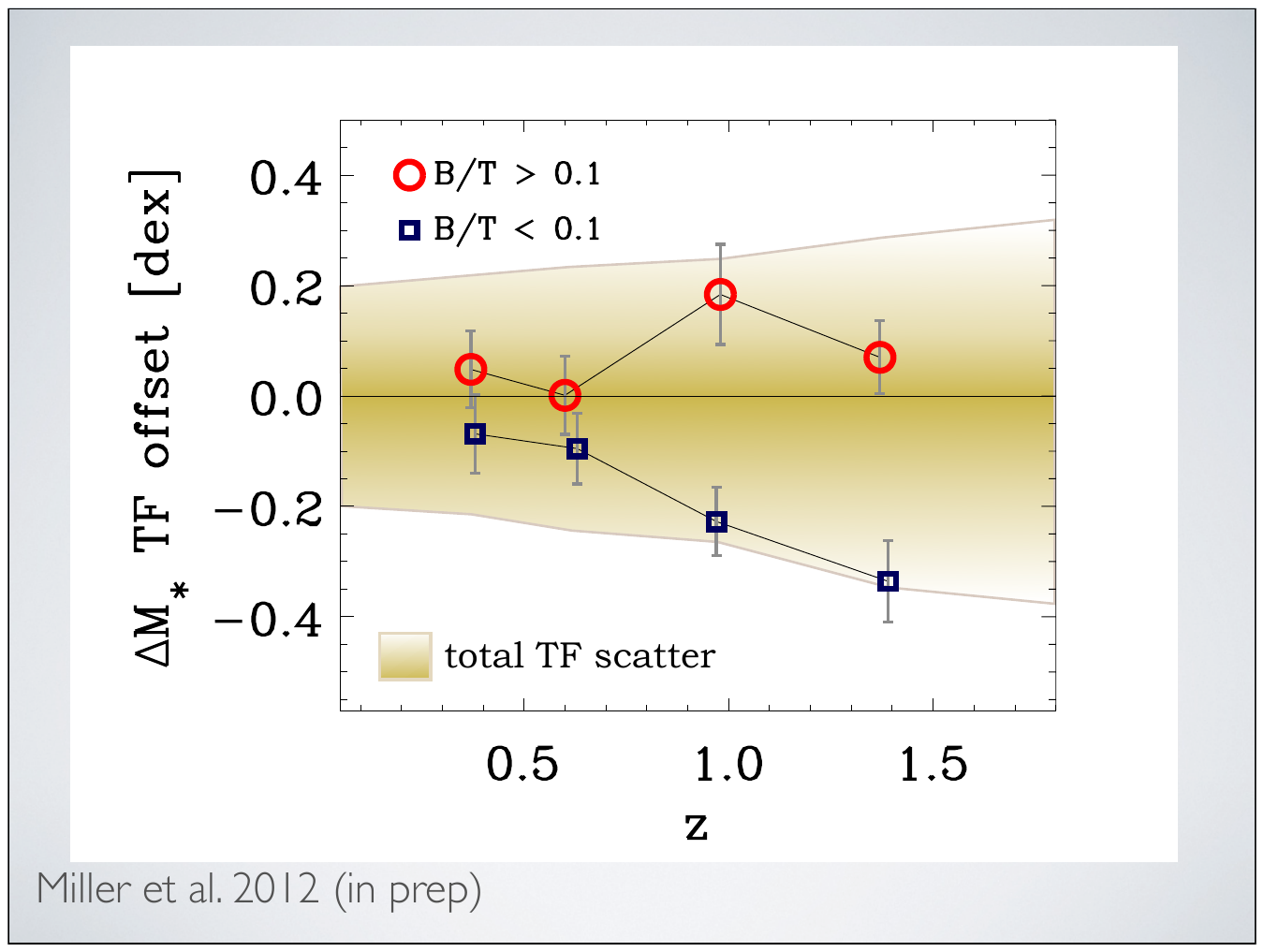}
      \caption{Evolution of the relative normalization of the $M_{\ast}$-TF relation, defined as the difference from the local relation (zero-points from fixed-slope fits). Bulgeless galaxies ($B/T$$<$0.1) are denoted separately from the bulge sample ($B/T$$>$0.1). We indicate total relation scatter with the gold gradient fill; the offset normalization of the two subsamples occurs within the total scatter of the $M_{\ast}$-TF relation. The bulgeless subsample becomes offset at high-$z$, but galaxies with bulges do not significantly deviate from the local relation (denoted by the solid black line, $\Delta$$M_{\ast}$-offset = 0.0).}
   \label{fig:fig2}
\end{figure}

We now examine the stellar mass Tully-Fisher ($M_{\ast}$-TF) relation
partitioned by morphology, in terms of the bulge/total ratio, $B/T$.
To facilitate this investigation, we separate our sample according to
the HST-derived \textsc{galfit} results into galaxies with prominent
bulges and those without (bulgeless disks and irregulars) as described
above. We plot the $M_{\ast}$-TF relation in four redshift bins
($0.2<z\leq0.5$,~$0.5<z\leq0.8$,~$0.8<z\leq1.2$,~$1.2<z\leq1.7$)
ensuring nearly equal sub-samples and look back time intervals (see
Fig. \ref{fig:bulgeless_tf}). Using the method described in M11, we
fit inverse linear regressions to each subsample and $z$-bin using a
fixed slope (of 3.70), the value of which was derived by fitting a
free slope to the entire sample. In the two highest redshift bins
($0.8<z\leq1.2$,~$1.2<z\leq1.7$), we see bulgeless disks are
significantly offset in the stellar mass (y-axis) normalization from
that of the local relation by $-0.23\pm0.06$ dex and $-0.34\pm0.07$
dex, respectively. In contrast, disks with significant bulges do not
deviate significantly from the local relation, nor in fact do
bulgeless disks in the two lower redshift bins (Fig. \ref{fig:fig2}).
The presence of a bulge appears to secure a disk galaxy on the
$M_{\ast}$-TF relation to within a scatter of 0.2 dex.
 
The question then arises as to whether increased scatter around the total $M_{\ast}$-TF relation from $z\simeq$1 to $z\simeq1.7$ can be accounted for largely via the inclusion of
bulgeless galaxies. In M12, the scatter from $z\sim1$ to $z\simeq1.7$ increased up to 60\%, whereas scatter across the 3 lower bins does not significantly evolve. 
Since the bulge-separated relations in the highest redshift bins have tighter relations separately than when both samples are combined, it seems likely that increased
scatter can be attributed to the zero-point shift of the bulgeless relation. The paucity of lower mass galaxies in the $0.8<z\leq1.2$ $z$-bin arising from the $K$-band magnitude
limit applied for the DEIMOS sample likely complicates this inference \citep[TF scatter increases to lower masses, e.g.,][]{begum2008}. We note that the LRIS sample 
forming the basis of the highest $z$-bin was not $K$-band limited.

To allow for various differences between sample sizes and distributions, we
quantify the offset significance for the bulgeless subsample using a
Monte Carlo `boot-strap' analysis. We fit two TF relations to
randomly-selected subsamples ($\times$10,000, with replacement) of our full data sample
(across all redshifts), ensuring subsamples of the same size (N) for
each $z$-bin (N=21/22, 34/15, 26/20, 21/12, for
bulgeless/bulge-dominated sets respectively).  With each pair of
randomly selected subsamples, we measure their relative normalization offset, and
fit Gaussians to the resulting histograms of the offset distributions (where the FWHM of each distribution is 0.082, 0.099, 0.086, 0.112 dex for each bin in increasing redshift). We also conduct an additional boot-strap analysis of N=10,000 where we select subsamples at random \emph{within} redshift bins (and also with replacement). This latter method, which accounts for redshift-dependence in the errors, results in broader normalization offset distributions (where FWHMs are 0.098, 0.110, 0.131, 0.132, respectively), and these distributions are plotted in the top left corner of each redshift-bin panel of Fig. \ref{fig:bulgeless_tf}. The true normalization offset
observed in each bin relative to the bootstrapped distribution of the latter method
yields offset significances of 
1.19$\sigma$,~0.88$\sigma$,~3.14$\sigma$,~3.07$\sigma$
for each bin in increasing redshift. Together this
translates to a confidence interval of greater than $99.8\%$ that the relation of the bulgeless
galaxies are offset at high-$z$ due to a genuine
effect  rather than random error or scatter. The slight decline in
significance in the highest bin is due to the reduced number of
galaxies with bulges in that bin, even though in real terms the
offset of the bulgeless relation is greatest in the highest $z$-bin
($-0.34$$\pm$0.07).

\section{Discussion}

In light of our results, we explore explanations of the offset co-evolution of stellar mass and total mass assembly in bulgeless galaxies from those which experienced earlier bulge growth. We first consider in \S~5.1 a favored picture in the literature of $z\sim2$ studies, and then turn in \S 5.2 and \S 5.3 to what may be a more self-consistent and simple picture for our results.
 
\subsection{Clump formation and migration} 
At $z$$\sim$2, low ratios of rotation-to-dispersion velocity support in disks \citep[$V$/$\sigma<2-5$,][]{forste2006, forste2009} have been interpreted in a simple picture whereby disks fragment and the resulting clumps migrate to the centers of galaxies to form bulges \citep{noguch1999,immeli2004,romeo2010}. 

Importantly, the velocity dispersion measured from such studies comes from the \emph{ionized} gas. The dispersion in the emission lines could be dominated by energy or momentum driven stellar feedback, rather than dispersion from dynamical pressure (as would be traced by the stars and cold gas). In our sample, while there is a spread in $V/\sigma$ values from 2--15 across all redshift bins ($90\%$ never rise above $V/\sigma$=10), no clear trend in $\sigma$ itself exists with respect to morphology or redshift. These values are more similar to the locally observed spread in $V/\sigma$ of ionized gas than those found in massive, star-forming galaxies at $z> 2$ \citep[$0<V/\sigma<5$,][]{genzel2008}. 

Furthermore, the clump migration picture does not explain why bulgeless galaxies arrive on the $M_{\ast}$-TF relation by intermediate redshift without forming bulges. In the clump migration picture, gas-rich bulgeless and barless disks would continue to suffer gravitational instabilities and clump formation until a bulge had formed and supposedly stabilized the disk. Some other mechanism is needed to explain the stabilization of bulgeless disks that have yet to form a substantial bulge/bar.

Additionally, central stellar velocity dispersions can support an increased stellar mass in the form of an accumulating bulge without significantly changing the rotational velocity of the surrounding disk. This would shift galaxies above the TF relation as their bulges grow without increasing rotational support, which is not supported by our results.

\subsection{An underestimation of gas masses in high-$z$ bulgeless disks}

A simpler explanation for our results is that bulgeless galaxies have higher gas fractions in their disks as compared to the rest of the sample. By adding total gas masses to our stellar masses, we may find the  baryonic Tully-Fisher is universal at all redshifts in rotationally-supported galaxies. To explore this possibility without direct gas mass measurements, we conduct the following exercise.

We compare the gas estimates from the empirically-based analytical method from M11 (Method 1), to the estimates based on the Kennicutt-Schmidt (K-S) relation (Method 2).  Described in detail in M11, Method 1 uses local gas fractions based on a galaxy's stellar mass and self-consistently adds the integrated specific star-formation rate from the galaxy's redshift (accounting for a gas recycling fraction based on the assumed Chabrier IMF). In Method 2, the K-S gas masses are determined from the rest-frame $B$-magnitude surface brightnesses, which are used to estimate a SFR density and thus a gas mass according to the inverted K-S relation \citep{kennic1998}.

This comparison reveals that the high redshift bulgeless disks have a significantly diminished B-band magnitude per area compared to the rest of the sample. That is to say Method 1 results in 1.11 times more gas than Method 2 with a 10\% scatter in bulgeless disks, whereas this factor is 1.43 in the rest of the sample.  If we increase the bulgeless disk gas mass estimates 30\% so that they align with the rest of the sample, then a universal baryonic TF relation is restored. 

Physically, a correction of this nature to our gas mass estimates suggests less metal-enriched gas in bulgeless disks, which is less efficient at forming stars. The relative youth of the bulgeless disks may be due to the lack of enriching outflows re-condensing at $z\sim1$ (but do by $z\sim0$). A lengthened `fountain' duty-cycle could reflect a shallower gravitational potential and lower star-formation surface densities \citep[i.e.,][etc.]{oppenh2008,finlat2008}, where outflows are slower to re-condense and metals more likely to escape in supernova-driven winds.

Also the simultaneous growth of supermassive black holes (SMBHs) at the centers of galaxies with their bulge mass is well known\citep{ferrar2000}. SMBHs are likely fed by gas that has sunk to the centers of galaxies via disk instabilities and/or mergers, a process which would similarly grow a bulge. It is then unsurprising that galaxies with no central bulge may have a much larger fraction of their gas still in their disks.

While the predictions of this exercise await testing via direct gas
mass observations, it does suggest interesting implications for a universal
baryonic-TF relation \citep[so far confirmed only locally,
e.g.][]{mcgaug2012}. Looking towards the future, these predictions can
be tested by determining the molecular and neutral hydrogen components
of these galaxies, via facilities such as the Atacama Large
Millimeter/submillimeter Array (ALMA), or future radio facilities
(e.g., MeerKAT, or ultimately the Square Kilometer Array, SKA).

\subsection{A $z$-dependent transition mass} 

Locally, low-mass bulgeless galaxies ($M_{\ast}<10^9M_{\odot}$) tend
to fall below the extrapolated $M_{\ast}$-TF relation from
$M_{\ast}>10^9M_{\odot}$ \citep{matthe1998,stark2009}. Since
this offset is similar to that seen in our high-redshift bulgeless
galaxies, it suggests that probing further down the stellar-mass
function at $0.2<z<0.8$ may uncover a similar transition to an offset of bulgeless disks at an
intermediate mass between that observed at $z\sim0$ and at$z>1$.

The physical significance of an evolving transition mass for bulgeless
galaxies could be understood via the `downsizing' concept
\citep[e.g.][]{cowie1996} or an evolving `mass floor' in galaxy
formation theory \citep[e.g.][]{bouche2010}.  These models seek to
explain why more massive galaxies formed earlier and faster than lower
mass galaxies, regardless of environment (appropriate for our study
since the role of environment would be subtle in our field sample).
These models are ultimately driven by the cosmic decline in accretion
rate, shutting down assembly of massive galaxies first by quickly
consuming their reservoirs.  A combination of the evolving ultraviolet
background with photo-ionizing radiation from the first stars could
create a transition mass, above which the cooling efficiency is
relatively higher, and below which a lack of self-shielding keeps
smaller, thinner disks from remaining neutral (keeping molecular gas
collapsing to form GMCs). Galaxies with bulges maintaining thicker
disks in thicker, steeper potential-wells could self-shield, and thus
form stars more efficiently than thin, bulgeless disks being adversely
affected by photo-ionizing radiation in shallower potential-wells.\\

In attempting to better understand drivers of disk assembly from our
results, we note a number of tensions regarding the picture where disks settle from the migration of large clumps into central bulges. Rather, a more self-consistent picture could be provided by an universal baryonic-TF relation, where better accounting of gas in high-$z$ disks could explain the offsets seen in tracking stellar mass with the total rotational support in galaxies. This may also predict a redshift-dependent transition mass which lowers with the age of the universe, below which bulgeless disks assemble their mass offset from the locally-defined $M_{\ast}$-TF relation, but not the baryonic-TF relation.

\acknowledgments

SHM thanks the Rhodes Trust and BFWG for supporting this work. MS acknowledges support from the Royal Society. We thank K.~Bundy for stellar mass estimates and spectral reduction, as well as A.~Newman for spectral reduction, and  helpful discussions with T. Treu.

{\it Facilities:} \facility{Keck I (LRIS)}, \facility{Keck II (DEIMOS)}, \facility{HST}.

\end{document}